\documentclass[aps,pre,reprint,groupedaddress]{revtex4-1}
\usepackage{graphicx}
\usepackage{amsmath}
\usepackage{array}
\usepackage{caption}
\usepackage{subcaption}
\usepackage{float}
\usepackage{placeins}
\DeclareMathOperator{\sech}{sech}

\begin{document}


\title{Spontaneous Quantum Teleportation in a Quenched Spin Lattice}


\author{N. Pilgram}
\author{T. D. Gutierrez}
\affiliation{Department of Physics, California Polytechnic State University, San Luis Obispo}

\date{\today}

\begin{abstract}
An  Ising-inspired numerical model is developed to study spontaneous quantum teleportation in a quenched spin lattice.  Quantum teleportation is an operation that can,  using entangled pairs of particles, transport a quantum state across arbitrary distances with high fidelity.  In doing so, it destroys the state in one location and relocates it to another.  In this context, teleportation serves as a long range interaction that randomly introduces correlations and disorder into a lattice.  In addition, different Bell state projection and entangled pair swapping models are also explored, as are the effects of decoherence.  The results are compared to the standard Ising model in one- and two-dimensions across several thermodynamic parameters versus temperature.
\end{abstract}

\pacs{}

\maketitle

\section{\label{sec:into}Introduction}

Quantum teleportation is an operation that can transport a pure quantum state across arbitrary distances with high fidelity~\cite{Qteleport}.  Normally, the process is performed actively by an experimenter~\cite{ExpQtel}\cite{SolidStateQtel}\cite{Exp Enganglement} and is a standard element of the growing arsenal of quantum information tools being developed for quantum computing, quantum cryptography, and other emerging quantum technologies~\cite{Qkey protocol}\cite{Qkey distribution}\cite{Qcomputing}.  Here, we explore the effects of spontaneous quantum teleportation using an Ising-inspired model in a quenched spin lattice.  That is, the teleportations occur without an active experimental intervention and the quantum correlation process is driven by the natural dynamics of a physical system in thermal equilibrium.  We refer to a quenched lattice as one which fixes the location of all particles and isolated spins in space. This is in contrast to an annealed lattice which would allow single spins and entangled particle pairs to diffuse and interact in the lattice versus time.

In addition to a lattice of fixed single-particles in $z$-component spin eigenstates used in the Ising model, all of our quantum correlation models require the presence of entangled pairs of particles known as Einstein-Podolsky-Rosen (EPR) pairs~\cite{EPR}.  In this context, the spin EPR pairs are treated quantum mechanically in the Bell basis.  These pairs can be considered to be correlated impurities in an otherwise thermalized spin system.  To introduce the EPR pairs into the spin system, and to emulate the measurement process, we consider several models of a quenched lattice and also consider decoherence effects. 

After initializing a randomized spin lattice, the fixed spins and EPR pairs are allowed to interact at some temperature.  Like the Ising model, these objects are subjected to a metropolis algorithm that allows their states to fluctuate based on nearest neighbor interactions.  However, now there are additional effective interactions that can spontaneously teleport single particle states, teleport entangled pairs, or generate entangled pairs, producing a variety of results.  

To isolate the effects of introducing entangled EPR pairs and spontaneous teleportation into the system, three cases are studied: pure Ising; Ising with teleportation; and teleportation alone.  The teleportation interaction includes pair swapping, where the teleported state is itself an EPR pair, as well as Bell state projections, which are local single particle measurements that project non-entangled pairs into entangled states.  The latter case has the effect of generating an EPR pair from two single-particle spins.  All of these spontaneous correlations alter the local spin interactions and can serve as a randomized long range interaction that introduces differing degrees of order and disorder, thus affecting various thermodynamic parameters.  These three models are compared with one another in both one and two dimensions.


The decoherence mechanism dilutes the purity of the Bell states, driving them into a mixed quantum configuration due to local interactions over a tunable characteristic time scale.  This has the effect of measuring pure two-particle entangled Bell states into mixed uncorrelated single particle states of definite spin.  In the limit of short decoherence times, the system approaches the expected Ising quenched lattice results.  In the limit of long decoherence times, the entangled pairs retain their quantum mechanical purity indefinitely.

After reviewing the mechanics of quantum teleportation, the Ising-inspired models are discussed in some detail for the one- and two-dimensional lattice.  The effects of teleportation on the system are isolated and studied across several thermodynamic observables varied quasi statically versus temperature such as: energy, specific heat, magnetization, and critical temperature.  In addition we study the entanglement density of the system.  Limitations of the models are discussed as well as possible future directions.







\section{\label{sec:Quantum}The Bell States, Quantum Teleportation, and Pair Swapping}

One remarkable result of quantum mechanics is the existence of entangled particles, particles who's quantum states are entangled, correlated, and dependent on one another. Once two particles are entangled, the quantum correlation and entanglement of states perpetuates over distance. For two spin one half particles, any possible entangled state in which they can reside can be described in terms of the Bell basis, a complete orthonormal basis for a set of two entangled spin one half particles. The four states, known as the Bell states, that comprise the Bell basis are
\begin{subequations}
\label{eq:Bellstates}
\begin{eqnarray}
\left|\Psi^{\left(-\right)}_{12}\right>&=\frac{1}{\sqrt{2}}\left(\left|\uparrow_1\downarrow_2\right>-\left|\downarrow_1\uparrow_2\right>\right)\label{eq:psi-}\\
\left|\Psi^{\left(+\right)}_{12}\right>&=\frac{1}{\sqrt{2}}\left(\left|\uparrow_1\downarrow_2\right>+\left|\downarrow_1\uparrow_2\right>\right)\label{eq:psi+}\\
\left|\Phi^{\left(-\right)}_{12}\right>&=\frac{1}{\sqrt{2}}\left(\left|\uparrow_1\uparrow_2\right>-\left|\downarrow_1\downarrow_2\right>\right)\label{eq:phi-}\\
\left|\Phi^{\left(+\right)}_{12}\right>&=\frac{1}{\sqrt{2}}\left(\left|\uparrow_1\uparrow_2\right>+\left|\downarrow_1\downarrow_2\right>\right)\label{eq:phi+}\\
\nonumber
\end{eqnarray}
\end{subequations}
where each subscript refers the respective particle~\cite{Qteleport}.

\begin{center}
\begin{figure}[!]
   \includegraphics[scale=.45]{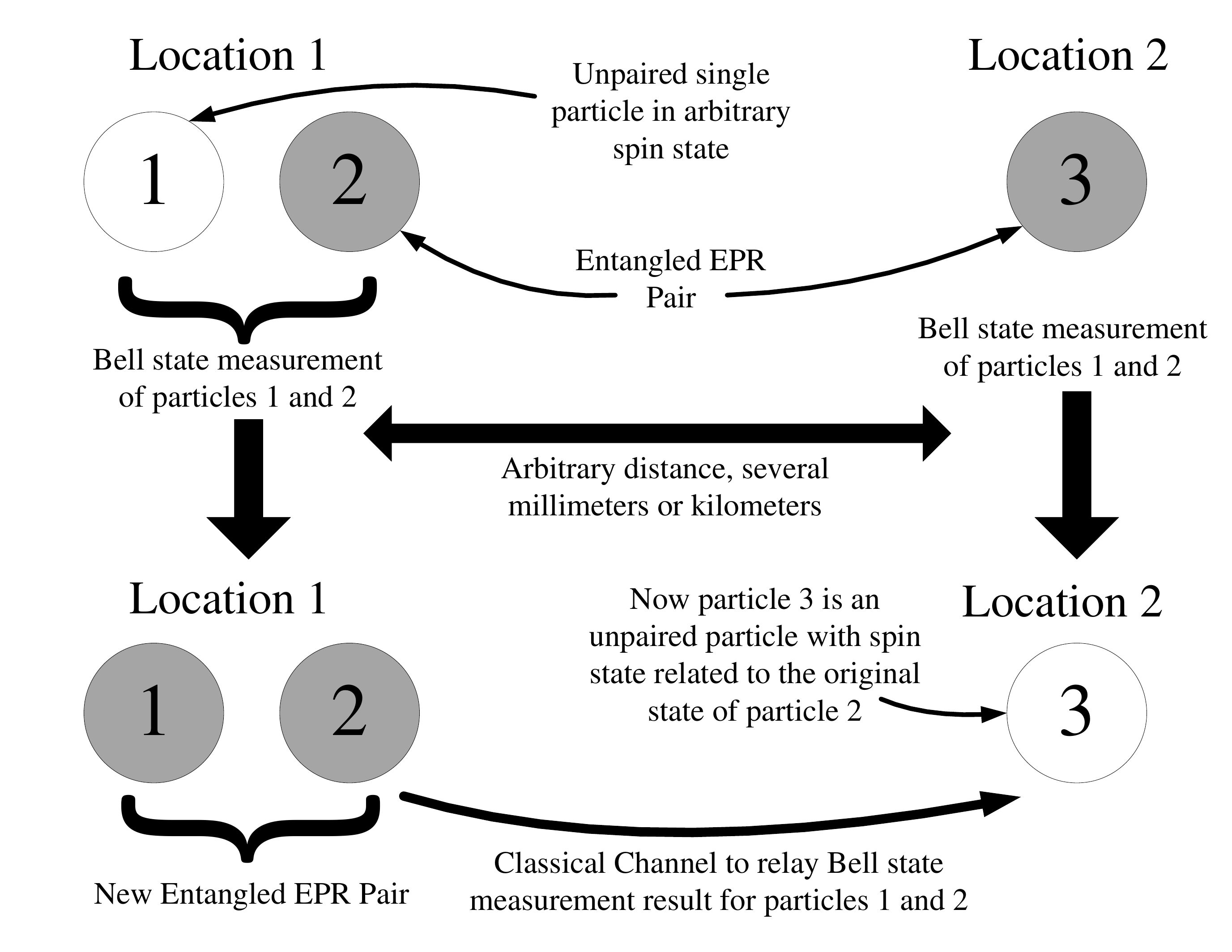} 
   \caption{Quantum teleportation schematic.Throughout the teleportation process, particles 1 and 2 are located separately from particle 3. Initially particles 2 and 3 are entangled in an EPR pair while particle 1 resides in the unknown state which is to be teleported. After a measurement of particles 1 and 2 is preformed in the Bell basis, particles 1 and 2 comprise a new EPR pair, and particle 3 resides in a non-entangled state related to the initial state of particle 1.To complete the teleportation process, the result of the Bell basis measurement of particles 1 and 2 is relayed to the reviver at Location 2, via a classical channel, allowing the receiver to determine which unitary operator to apply to completely reconstruct the initial state of particle 1.\label{fig:Qtelf1}}
\end{figure}
\end{center}

The non-local effects of quantum entanglement can be utilized to transmit quantum states, or quantum information, over distances without physically transferring any particles themselves. This transmission process is known as quantum teleportation. The nature of entangled particles and the quantum teleportation process also does not require the sender to have any knowledge of the transmitted state or of the location of the receiver~\cite{Qteleport}. Simple quantum teleportation can be accomplished using a three particle system. This quantum teleportation process is outlined in Fig.~\ref{fig:Qtelf1}. The first particle, particle 1, resides in the unknown state which is to be teleported, while the other two particles, particles 2 and 3, are prepared in a Bell state. The unknown state, particle 1, can be described by
\[\left|\phi_1\right>=a\left|\uparrow_1\right>+b\left|\downarrow_1\right>\]
where $a$ and $b$ can be complex and satisfy the normalization condition $\left|a\right|^2+\left|b\right|^2=1$. In order to transmit the unknown state, $\left|\phi_1\right>$,  of particle 1 to another location, particle 1 and one of the entangled particles, particle 2, must be separated from the other entangled particle, particle 3, as shown in Fig.~\ref{fig:Qtelf1}. The location of particle 3 will be the location where the quantum state of particle 1 will be transmitted. To initiate the quantum teleportation process, a measurement of particles 1 and 2 must be preformed in the Bell operator basis. A measurement of particles 1 and 2 in the Bell operator basis will entangle particles 1 and 2 into one of the four Bell states given by Eqn.~\ref{eq:Bellstates}. Since particles 2 and 3 where previously correlated, the resulting state of particle 3, which is in another location, is dependent on which Bell state particles 1 and 2 are projected into by the measurement. If particles 2 and 3 initially resided in the singlet state, given by Eqn.~\ref{eq:psi-}, then expressing the wave function of the entire system in terms of the Bell basis of particles 1 and 2, as is done in ~\cite{Qteleport}, reveals that the four possible resulting states of particle 3 are
\begin{eqnarray*}
\left|\phi_3\right>_1&=&-a\left|\uparrow_3\right>-b\left|\downarrow_3\right>\\
\left|\phi_3\right>_2&=&-a\left|\uparrow_3\right>+b\left|\downarrow_3\right>\\
\left|\phi_3\right>_3&=&a\left|\downarrow_3\right>+b\left|\uparrow_3\right>\\
\left|\phi_3\right>_4&=&a\left|\downarrow_3\right>-b\left|\uparrow_3\right>
\end{eqnarray*}

\onecolumngrid
\twocolumngrid

The resulting $\left|\phi_3\right>_1$ state corresponds to a simple phase shift of the original unknown state, $\left|\phi_1\right>$, while the other three correspond to 180$^{\circ}$ rotations of $\left|\phi_1\right>$ about the x, y, and z axes. Since the resulting state of particle 3 is correlated to the Bell basis measurement result of particles 1 and 2, the measurement result can be relayed to the location of particle 3, indicating which unitary operator (if any) the receiver must apply to particle 3 to completely reconstruct the original unknown state $\left|\phi_1\right>$ and complete the teleportation process~\cite{Qteleport}. The process of quantum teleportation is more rigorously described in ~\cite{Qteleport}.

A special case of quantum teleportation occurs when all particles in the system reside in entangled states. Consider a system of four entangled particles, where particles 1 and 2 are entangled in a Bell state and particles 3 and 4 are entangled in a separate Bell state. 
If particles 2 and 3 are then entangled and projected into a Bell state, it follows that the particles with which they were previously paired, particles 1 and 4, will also be entangled and projected into a Bell state. It turns out that no matter what combination of Bell states particles 1 and 2 and particles 3 and 4 reside in, if particles 2 and 3 are projected into a bell state the final state of the system will be one of the following four,
\begin{subequations}
\label{eq:pair swap result}
\begin{eqnarray}
\left|\Psi_{1234}\right>&=\left|\Phi^{\left(+\right)}_{23}\right>\otimes\left|\Phi^{\left(+\right)}_{14}\right>\label{eq:pair swap1}\\
\left|\Psi_{1234}\right>&=\left|\Phi^{\left(-\right)}_{23}\right>\otimes\left|\Phi^{\left(-\right)}_{14}\right>\label{eq:pair swap2}\\
\left|\Psi_{1234}\right>&=\left|\Psi^{\left(+\right)}_{23}\right>\otimes\left|\Psi^{\left(+\right)}_{14}\right>\label{eq:pair swap3}\\
\left|\Psi_{1234}\right>&=\left|\Psi^{\left(-\right)}_{23}\right>\otimes\left|\Psi^{\left(-\right)}_{14}\right>.\label{eq:pair swap4}
\end{eqnarray}
\end{subequations}
where the resulting entangled states of particles 2 and 3 and particles 1 and 4 are the Bell states given in Eqn.~\ref{eq:Bellstates}~\cite{PairSwap}. Thus, Eqn.~\ref{eq:pair swap result} indicates that whichever Bell state particles 2 and 3 are projected into is the same Bell state particles 1 and 4 are projected into. This creation of two new EPR pairs causes all four particles in the system to swap entanglement partners and is therefore refereed to as pair, or entanglement, swapping  The model discussed in this paper, investigates the thermodynamical effects of a Bell state projection interaction in a quenched spin lattice. A Bell state projection interaction or measurement is responsible for both quantum teleportation and pair swapping as well as the creation of entangled EPR pairs.

Normally, the quantum teleportation process requires a classical channel to complete the operation with 100\% fidelity.  That is, the appropriate 180$^{\circ}$ rotation required to finish the process with complete certainty needs to be communicated to the member of the EPR pair prepared to receive the unknown state.  In our case, we dispense with the classical channel and allow statistics to determine how the unknown state is transported across the lattice.  As described above, in the standard treatment, the unknown state has complex coefficients $a$ and $b$ which characterize the state to be teleported.  In our case, to simplify the model for this treatment, $a$ and $b$ are either 0 or 1.  That is, the unknown states are always $z$-component spin eigenstates.  As a result, up to an overall phase that does not affect the dynamics of the model, the unknown quantum state is always teleported with a 50\% fidelity.  Otherwise, it is the opposite spin state is teleported.  What makes this interesting, regardless of the fidelity, is that a local interaction in one part of the lattice can force the spin state into another part of the lattice.  A random spin state is inserted into the lattice even if it would not be normally energetically favorable to do so.

Future models will incorporate a lattice of spin states with arbitrary complex coefficients.  This is equivalent to a lattice of Bloch spheres.  By retaining all the phase information, a richer variety of interactions can be explored.  Decoherence could then be modeled using a density matrix approach whereby the state vectors not only reside on the surface of the Bloch spheres, but can diffuse into the bulk of the spheres in time due to interactions.  


\section{\label{sec:Ising Model}The Ising Model of a Ferromagnet and the Metropolis Algorithm}

The Ising model of a ferromagnet is a simplified model which provides relevant insight into the thermodynamic behavior of a ferromagnet, specifically the spontaneous phase change from a non-magnetic to magnetic state. The Ising model of a ferromagnet consists of a set of spins, and thus magnetic moments, arranged in a regular lattice, usually linear, square, or cubic depending on dimension. These spins can take one of two values +1 (spin up) or -1 (spin down). Each spin in the lattice is acted upon by its immediate neighbors, where the force is dependent on the relative orientation of the neighboring spins. Aligned spins are favored while anti-aligned spins are not. Since each spin has a magnetic moment, all spins can be acted upon  by an external magnetic field as well. The total energy of the system is given by
\begin{equation}
\label{eq:Tot Ising Energy}
E=-\epsilon\sum_{(i,j)}{s_is_j}-mB_{\rm{ext}}\sum_i{s_i}
\end{equation}
where $s_i$ is the spin state of the ith spin, $\epsilon$ is the magnitude of the energy of two neighboring spins, $m$ is the magnetic moment of a spin, $B_{\rm{ext}}$ is the external magnetic field, and the sum over $(i,j)$ indicates a sum over all neighboring spins~\cite{IsingKramers}. The first term in Equation~\ref{eq:Tot Ising Energy} represents the energy from the interaction between neighboring spins, while the second term represents the energy from each spins interaction with an external magnetic field. Therefore, in the absence of an external magnetic field the energy of the system is
\begin{equation}
\label{eq:Ising Energy}
E=-\epsilon\sum_{(i,j)}{s_is_j}
\end{equation}
where the sum is agin carried over all neighboring spins. The total magnetization of the system is given by
\begin{equation}
\label{eq:Ising Magnetization}
M=m\sum_i{s_i}
\end{equation}
where in this case the sum is carried over all spins. The Ising model holds for any dimension, however the one dimensional Ising model does not result in a phase change. The two and three dimensional Ising models do display phase changes. Several analytical solutions to the two dimensional Ising model have been completed, two of which can be found in ~\cite{IsingKramers} ~\cite{IsingOnsager}, however no analytical solution to the three dimensional Ising model has been found. 

In addition to finding analytical solutions, the Ising model can also be investigated through Monte Carlo algorithms. However, since a lattice of any considerable size has an innumerable number of possible states, it is practical to utilize a Monte Carlo algorithm with importance sampling, specifically the Metropolis algorithm~\cite{Metropolis}. Under the Metropolis algorithm, new random state configurations are generated based on the Boltzmann probability
\begin{equation}
\label{eq:Metropolis Prob} 
e^{-\Delta E/k_BT}
\end{equation}
where $\Delta E$ is the energy difference, $k_B$ is Boltzmann's constant, and $T$ is the temperature of the system.

For the Ising Model, the Metropolis algorithm is specifically executed as follows. A random spin within the lattice is chosen and the energy difference, $\Delta E$, that would result from flipping the spin is calculated. If $\Delta E \leq 0$, i.e., flipping the spin lowers the energy of the system, then the spin is flipped. However, if $\Delta E > 0$, then the spin is flipped with the probability given in Eqn.~\ref{eq:Metropolis Prob}. Then the thermodynamical quantities, such as energy, for this new configuration can be calculated. The average of a thermodynamical quantity over all configurations generated at a specific temperature results in that thermodynamical quantities value for that temperature.  The model discussed in this paper is based on the Ising model and is therefore investigated under the Metropolis algorithm. For comparative purposes this paper presents the results of the Ising model in addition to the modified case.

\section{\label{sec:Qteleport Model}Ising-Inspired model of Quantum Teleportation in a Spin Lattice}

In order to investigate the thermodynamical effects of quantum teleportation in a quenched spin lattice, an Ising-inspired model was developed. By directly modifying the Ising model to encompass entangled Bell states as well as Bell state projection and quantum decoherence interactions, the quantum teleportation of spin states within the lattice could occur. The model is structured as a discrete lattice of stationary particles, just as the Ising model. However, unlike the Ising model, each particle can reside in one of three distinct states, spin up, spin down, or an entangled Bell state. Just as in the Ising model, the spin up state is represented by $s=+1$ and the spin down state is represented by $s=-1$. Each particle in an entangled Bell state resides in a perfect $50/50$ superposition of a spin up and spin down state, thus the energy between a particle in a Bell state and any neighboring particle is the average of the energies given if the Bell state particle was in either a spin up or spin down state. Since the average of these two energies is zero, the energy between a particle in a Bell state and any neighboring particle is zero. Therefore, any particle residing in a Bell state can be represented by $s=0$. 

The total energy of the lattice is still defined as it was for the Ising model and is given by Eqn.~\ref{eq:Ising Energy}, however the possible values of $s_i$ are now +1, -1, and 0 instead of simply +1 or -1. Similarly, the magnetization of the lattice is still defined by Eqn.~\ref{eq:Ising Magnetization} with the additional $s_i$ value of zero. In addition to adding Bell states, a Bell state projection interaction executed via the Metropolis algorithm and a time dependent quantum decoherence interaction were included. Since both of these interactions are dependent of the specific Bell state in which an EPR pair resides, the specific Bell state of all EPR pairs is also determined and tracked throughout the model.

\subsection{\label{sec:Bell State}Bell State Projection Interaction}

The Bell state projection interaction projects two particles within the lattice into an entangled Bell state. The projection of two particles into a Bell state is the mechanism that is responsible for both quantum teleportation and pair swapping as discussed in Section~\ref{sec:Quantum}. Therefore, the Bell state projection interaction may result in the long range teleportation of quantum states within the lattice or in pair swapping. Just as with the basic spin flip interaction in the Ising model, the Bell state projection interaction is executed under the Metropolis algorithm. Under the Metropolis algorithm, the Bell state projection interaction is executed as follows. Two random adjacent particles within the lattice are chosen and the energy difference, $\Delta E$, that would result from projecting the two particles into a Bell state is calculated. For the case in which the teleportation of a quantum state would result from the interaction, the energy change due to the teleportation of the quantum state is not considered in $\Delta E$, since the exact state teleported is random. Also, it is important to note that $\Delta E=0$ in the case of a pair swap because all particles will remain in Bell states, i.e., $s=0$. If $\Delta E\leq 0$, then the two particles are projected into a Bell state, where the specific Bell state into which the two particles are projected is random and each Bell state is equally probable. If $\Delta E >0$ then the particles are randomly projected into a Bell state based on the probability given in Equation~\ref{eq:Metropolis Prob}. Again, the specific Bell state into which the particles are projected is random where each Bell state is equally likely. When two particles are projected into a Bell state there are three possible results, where each result depends on the initial states of the involved particles. Each of the three possible resulting cases is outlined below. 

\begin{center}
\begin{figure}[!]
  \includegraphics[scale=.45]{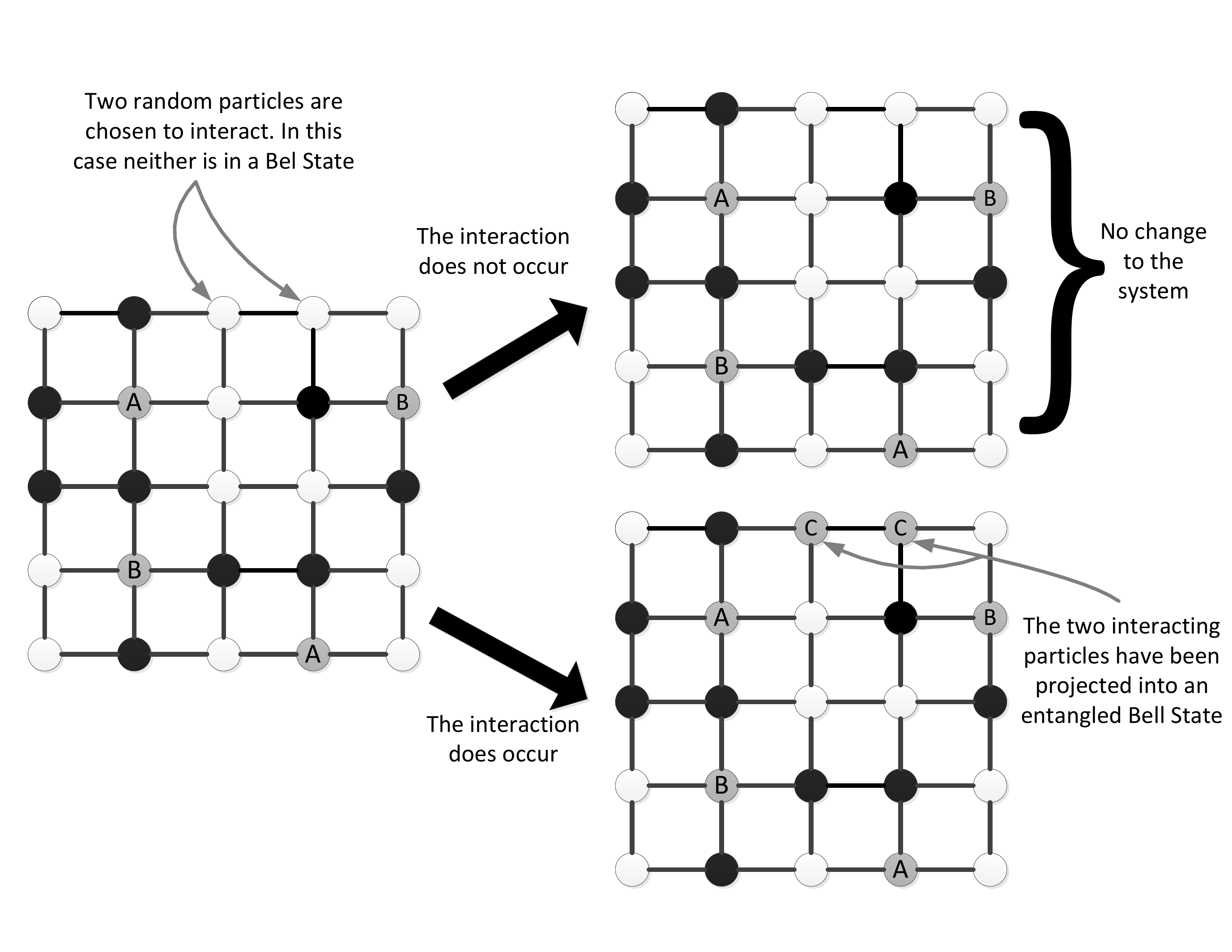} 
   \caption{Bell state projection schematic in a $5\times5$ lattice. White represents spin up, black represents spin down, and gray represents a Bell state. Each EPR pair is separately labeled, so that the particles labeled A represent one EPR pair while the particles labeled B and C represent other separate EPR pairs. \label{fig:intf3}}
\end{figure}
\end{center}

The first and simplest case, case 1, is the projection of two non-entangled particles into an entangled Bell state. This occurs if both particles involved in the Bell state projection interaction reside in either a spin up or spin down state. A visual representation of a simple Bell state projection is show in Fig.~\ref{fig:intf3}.

\begin{center}
\begin{figure}[!]
   \includegraphics[scale=.45]{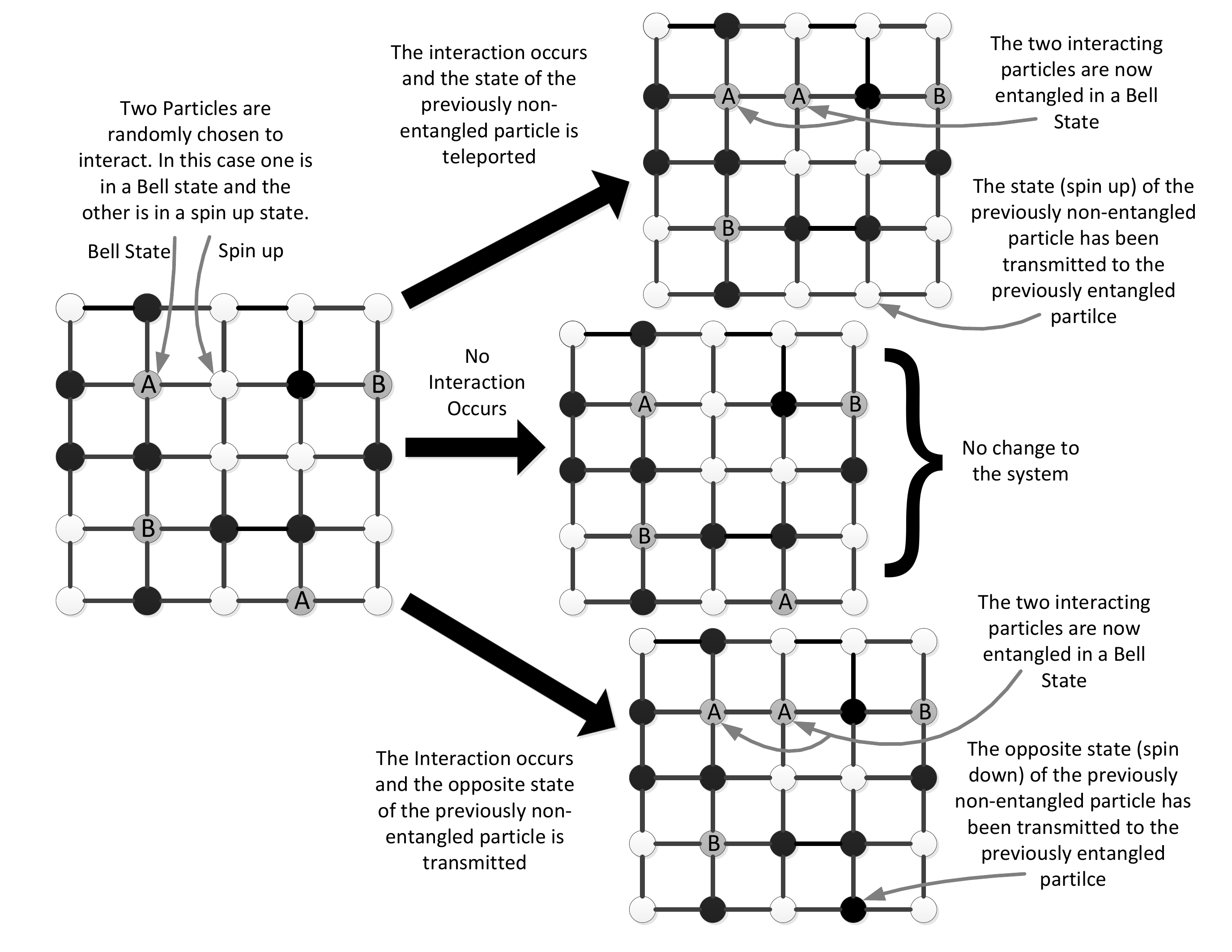} 
   \caption{Quantum teleportation schematic in a $5\times5$ lattice. White represents spin up, black represents spin down, and gray represents a Bell state. Each EPR pair is separately labeled so that the particles labeled A represent one EPR pair while the particles labeled B represent another separate EPR pair.\label{fig:intf1}}
\end{figure}
\end{center}

The second and most interesting case, case 2, is the teleportation of a quantum state. This occurs when one particle involved in the interaction resides in either a spin up or spin down state and the other resides in one of the four Bell states. Considering all possible combinations of the two spin states and the four Bell states results in eight possible situations described by the following product state wave functions,
\begin{widetext}
\begin{subequations}
\label{eq:Qtel results}
\begin{eqnarray}
\left|\Psi^{\left(-\right)}_{23}\right>\left|\uparrow_1\right>&=&\frac{1}{2} \left[-\left|\Psi^{\left(-\right)}_{12}\right>\left|\uparrow_3\right>-\left|\Psi^{\left(+\right)}_{12}\right>\left|\uparrow_3\right>+\left|\Phi^{\left(-\right)}_{12}\right>\left|\downarrow_3\right>+\left|\Phi^{\left(+\right)}_{12}\right>\left|\downarrow_3\right>\right]\\
\left|\Psi^{\left(-\right)}_{23}\right>\left|\downarrow_1\right>&=&\frac{1}{2} \left[-\left|\Psi^{\left(-\right)}_{12}\right>\left|\downarrow_3\right>-\left|\Psi^{\left(+\right)}_{12}\right>\left|\downarrow_3\right>+\left|
\Phi^{\left(-\right)}_{12}\right>\left|\uparrow_3\right>+\left|\Phi^{\left(+\right)}_{12}\right>\left|\uparrow_3\right>\right]\\
\left|\Psi^{\left(+\right)}_{23}\right>\left|\uparrow_1\right>&=&\frac{1}{2} \left[\left|\Psi^{\left(-\right)}_{12}\right>\left|\uparrow_3\right>+\left|\Psi^{\left(+\right)}_{12}\right>\left|\uparrow_3\right>+\left|\Phi^{\left(-\right)}_{12}\right>\left|\downarrow_3\right>+\left|\Phi^{\left(+\right)}_{12}\right>\left|\downarrow_3\right>\right]\\
\left|\Psi^{\left(+\right)}_{23}\right>\left|\downarrow_1\right>&=&\frac{1}{2} \left[-\left|\Psi^{\left(-\right)}_{12}\right>\left|\downarrow_3\right>+\left|\Psi^{\left(+\right)}_{12}\right>\left|\downarrow_3\right>-\left|\Phi^{\left(-\right)}_{12}\right>\left|\uparrow_3\right>+\left|\Phi^{\left(+\right)}_{12}\right>\left|\uparrow_3\right>\right]\\
\left|\Phi^{\left(-\right)}_{23}\right>\left|\uparrow_1\right>&=&\frac{1}{2} \left[-\left|\Psi^{\left(-\right)}_{12}\right>\left|\downarrow_3\right>-\left|\Psi^{\left(+\right)}_{12}\right>\left|\downarrow_3\right>+\left|\Phi^{\left(-\right)}_{12}\right>\left|\uparrow_3\right>+\left|\Phi^{\left(+\right)}_{12}\right>\left|\uparrow_3\right>\right]\\
\left|\Phi^{\left(-\right)}_{23}\right>\left|\downarrow_1\right>&=&\frac{1}{2} \left[-\left|\Psi^{\left(-\right)}_{12}\right>\left|\uparrow_3\right>+\left|\Psi^{\left(+\right)}_{12}\right>\left|\uparrow_3\right>+\left|\Phi^{\left(-\right)}_{12}\right>\left|\downarrow_3\right>-\left|\Phi^{\left(+\right)}_{12}\right>\left|\downarrow_3\right>\right]\\
\left|\Phi^{\left(+\right)}_{23}\right>\left|\uparrow_1\right>&=&\frac{1}{2} \left[\left|\Psi^{\left(-\right)}_{12}\right>\left|\downarrow_3\right>+\left|\Psi^{\left(+\right)}_{12}\right>\left|\downarrow_3\right>+\left|\Phi^{\left(-\right)}_{12}\right>\left|\uparrow_3\right>+\left|\Phi^{\left(+\right)}_{12}\right>\left|\uparrow_3\right>\right]\\
\left|\Phi^{\left(+\right)}_{23}\right>\left|\downarrow_1\right>&=&\frac{1}{2} \left[-\left|\Psi^{\left(-\right)}_{12}\right>\left|\uparrow_3\right>+\left|\Psi^{\left(+\right)}_{12}\right>\left|\uparrow_3\right>-\left|\Phi^{\left(-\right)}_{12}\right>\left|\downarrow_3\right>+\left|\Phi^{\left(+\right)}_{12}\right>\left|\downarrow_3\right>\right]\\
\nonumber
\end{eqnarray}
\end{subequations}
\end{widetext}
where particles 1 and 2 are involved in the Bell state projection interaction, particle 1 initially resides in either a spin up or spin down state, and particles 2 and 3 are initially entangled in one of the four Bell states. Equation~\ref{eq:Qtel results} shows that every spin state, Bell state combination results an 50$\%$ probability that the teleported state, final state of particle 3, will be spin up and a 50$\%$ probability that the teleported state will be spin down. Therefore, since the specific spin state that is teleported is random, the teleported state may  actually increase the energy of the gas even if the projection of particles 1 and 2 into a Bell state lowers the energy. Conversely, the teleported state can also decrease the energy of the gas even though the projections of particles 1 and 2 into a Bell state raises the energy. A visual representation of a teleportation is shown in Fig.~\ref{fig:intf1}.



\begin{center}
\begin{figure}[!]
   \includegraphics[scale=.45]{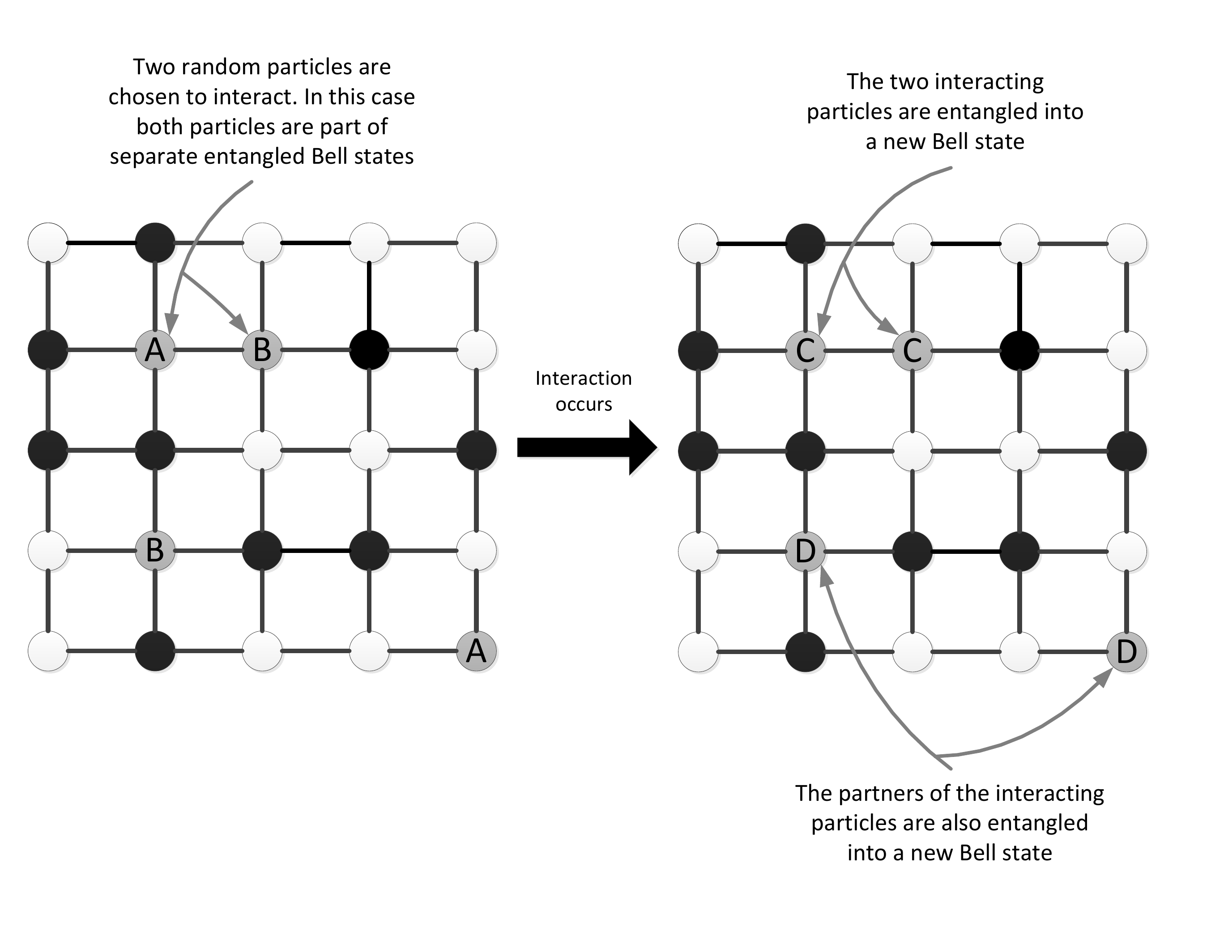} 
   \caption{Pair swap schematic in a $5\times5$ lattice. White represents spin up, black represents spin down, and gray represents a Bell state. Each EPR pair is separately labeled  so that the particles labeled A represent one EPR pair while the particles labeled B, C, and D represent other separate EPR pairs. After the interaction, the particles labeled C and D are the same particles previously labeled A and B, however, they are labeled differently because they are now entangled in new EPR pairs with different partners.\label{fig:intf2}}
\end{figure}
\end{center}

The third and final case, case 3, is that of a pair swap. This occurs when both particles involved in the Bell state projection interaction are part of separate entangled Bell states. The pair swap will occur in the manner described in Section~\ref{sec:Quantum} and the four possible, and equally likely, resulting wavefunctions are given by Eqn.~\ref{eq:pair swap result}. Since $\Delta E=0$ for all pair swap cases, a pair swap will always occur under the Metropolis algorithm. The visual representation of a pair swap is shown in Fig.~\ref{fig:intf2}.

Finally, it is important to note the inherent tendency of the Bell state projection interaction to project all particles in the lattice into Bell states. Since the interaction only projects particles into Bell states, as time and iteration number progress, the number of particles in Bell states increases. Thus, after many iterations most, if not all, particles will reside in a Bell state. This will result in a near zero energy. However, this effect can me mitigated by a quantum decoherence interaction, which will return particles in Bell states to the more classical spin states.


\onecolumngrid
\twocolumngrid

\onecolumngrid
\twocolumngrid

\onecolumngrid
\twocolumngrid

\FloatBarrier

\subsection{\label{sec:Decoherence}Quantum Decoherence Interaction}

The quantum decoherence interaction provides a very simplified model and execution of quantum decoherence. Quantum decoherence~\cite{Decoherence} is the process by which a quantum system looses its quantum coherence and devolves into a semi-classical or classical state. The loss of quantum information which causes the quantum system to devolve is a result of the interaction between the quantum system itself and the environment, which is also treated as a quantum mechanical system. Quantum correlations between the quantum system and the environment allow quantum information to be dispersed throughout the environment. This dispersion of quantum information increases the entropy of the system. 

\begin{center}
\begin{figure}[!]
   \includegraphics[scale=.6]{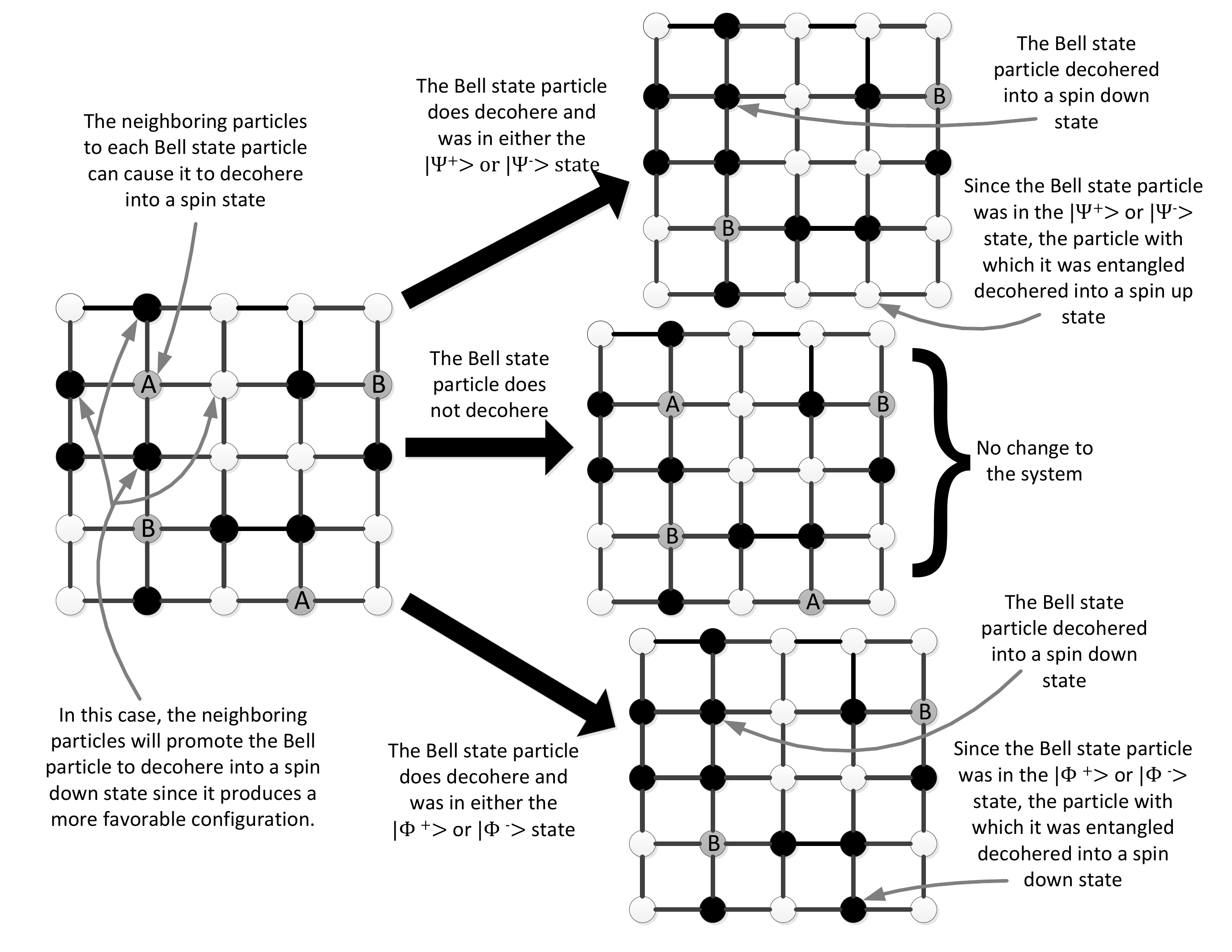} 
   \caption{Decoherence schematic in a $5\times5$ lattice. White represents spin up, black represents spin down, and gray represents a Bell state. Each EPR pair is separately labeled so that the particles labeled A represent one EPR pair while the particles labeled B represent another separate EPR pair. When a particle in an EPR pair decoheres, its partner will also decohere in a correlated manner consistent with the Bell state in which the particles used to reside. \label{fig:intf4}}
\end{figure}
\end{center}

The quantum decoherence interaction simulates a simplified interaction between a particle in a Bell state and its neighboring particles, i.e, local environment. The interaction between a particle in Bell state and its neighbors will cause the particle to disperse quantum information and decohere into the more classical spin up and spin down states. The dispersion of quantum information is a result of the neighboring particles preforming a measurement on the Bell state particle in a pointer basis determined by the orientations of the neighboring particle's spins. For example, if a majority of the neighboring particles of a Bell state are spin up, then the Bell state particle decohering into a spin up state will be energetically favorable. Thus, the neighboring particles will "perform" an effective measurement on the Bell state particle causing it to decohere into a spin up state. A similar decoherence of the Bell state particle to a spin down state would occur if the majority of the neighboring particles were in spin down states. The random likelihood that a Bell state will decohere operates on a time dependent probability, where the probability increases with the time the particle has resided in a Bell state. The decoherence probability is given by
\begin{equation}
\mathcal{P}_d=1-e^{-t/\tau}
\label{eq:decohereProb}
\end{equation}
where $t$ is the time in which the particle has been in a Bell state, and $\tau$ is the characteristic decoherence time. A visual representation of the quantum decoherence interaction is shown in Fig.~\ref{fig:intf4}.


When a particle in an EPR pair decoheres, its partner will also decohere in a correlated manner. This correlated decoherence will occur in a manner consistent with the Bell state in which the two particles were previously entangled. For example, if a particle residing in either the $\left|\Psi^{\left(+\right)}_{12}\right>$ or $\left|\Psi^{\left(-\right)}_{12}\right>$ state decoheres into a spin up state, its partner will decohere into a spin down state. In the general case, when a particle in either the $\left|\Psi^{\left(+\right)}_{12}\right>$ or $\left|\Psi^{\left(-\right)}_{12}\right>$ state decoheres into a spin state, its partner will decohere into the opposite spin state. Conversely, when a particle in either the $\left|\Phi^{\left(+\right)}_{12}\right>$ or $\left|\Phi^{\left(-\right)}_{12}\right>$ state decoheres into a spin state, its partner will decohere into the same spin state. In addition to the quantum decoherence interaction modeling the quantum mechanical interaction between a particle in a Bell state and its local environment, the interaction also mitigates the Bell state projection interaction's inherent tendency to project all particles in the system into Bell states.



\section{Numerical Details\label{sec:Num Details}}
To isolate and understand the thermodynamical effect of a Bell state projection interaction in a quenched spin system, six different interaction models were conducted under the Metropolis algorithm, three in one-dimension and three in two-dimensions. The one-dimensional models and their designations are as follows; the pure one-dimensional Ising model (1D Ising Model), the combination of the Bell state projection interaction and the quantum decoherene interaction (Model 1A), and the combination of the Ising spin flip interaction, Bell state projection interaction, and quantum decoherence interaction (Model 1B). The corresponding two-dimensional models are as follows; The pure two-dimensional Ising model (2D Ising Model), the combination fo the Bell state projection and quantum decoherence interactions (Model 2A), and the combination of the Ising spin flip, Bell state projection, and quantum decoherence interactions (Model 2B). A summery of all models by name, dimensionality, and comprising interactions  is given in Table~\ref{table: Models}.

\onecolumngrid
\twocolumngrid

\onecolumngrid
\begin{widetext}
\begin{table*}[h]
\caption{Summery of models with name, dimensionality, and comprising interactions.\label{table: Models}}
\begin{ruledtabular}
\begin{tabular}{|c|c|c|c|c|}
Model & Dimension & Ising spin flip interaction & Bell state projection interaction & Quantum decoherence interaction\\
\hline
1D Ising Model & 1 & yes & no & no\\
Model 1A & 1 & no & yes & yes\\
Model 1B & 1 & yes & yes & yes\\
2D Ising Model & 2 & yes & no & no\\
Model 2A & 2 & no & yes & yes\\
Model 2B & 2 & yes & yes & yes\\
\end{tabular}
\end{ruledtabular}
\end{table*}
\end{widetext}
\twocolumngrid

The three one-dimensional models were studied on a $1\times40$ linear lattice and the two-dimensional models were studied on a $40\times40$ square lattice. The $1\times40$ and $40\times40$ lattices provided the optimal combination of of data variation reduction and computation time. The various models were iterated over varying temperature ranges, which were dependent on the dimensionality and $\tau$ value used. Each temperature range was split into a finite number of evenly spaced points so that a data point density of about 306 data points per temperature [$k_BT/\epsilon$] resulted, where $\epsilon$ is the neighboring spin interaction energy given in  Eqn.~\ref{eq:Ising Energy}. The Metropolis algorithm was executed 50,000 time at each temperature point. The resulting lattice for each iteration and temperature point was used as the initial lattice for the next. To determine the effects of varying $\tau$ values, given in Eqn.~\ref{eq:decohereProb}, all interaction models involving the Bell state projection and quantum decoherence interactions (Model 1A, Model 1B, Model 2A, and Model 2B) were conducted with $\tau$ values of $10^{-7}I_{\rm{tot}}$,  $10^{-5}I_{\rm{tot}}$,  $10^{-4}I_{\rm{tot}}$,  $10^{-3}I_{\rm{tot}}$,  $10^{-2}I_{\rm{tot}}$,  $10^{-1}I_{\rm{tot}}$, $I_{\rm{tot}}$, and $10^{2}I_{\rm{tot}}$, where $I_{\rm{tot}}$ is the total number of Iterations over all temperature steps. The temperature dependent thermodynamical quantities of energy, magnetization, specific heat, and entanglement density were determined from each execution.

\section{Preliminary Results\label{sec:Results}}
The preliminary thermodynamic results in both one- and two-dimensions displayed critical behavior varying from that displayed by the Ising model. For both one- and two-dimensions the existence and location of a critical temperature was dependent on the $\tau$ parameter used in the decoherence interaction. For small $\tau$ values, which correlates to short decoherence times, the one-dimensional results displayed no critical behavior. Thus the low $\tau$ value one-dimensional energies and specific heats, as well as those obtained for the one-dimensional Ising model, were fitted to the respective one-dimensional Ising model analytical energy and specific heat. The one-dimensional Ising model results matched the analytical functions almost exactly, however the other low $\tau$ value results displayed some variation. Where critical behavior occurred, the magnetization and specific heat data was fitted to the respective proportional power laws of
\begin{equation}
\label{eq:M power law}
M\propto\left|T_c-T\right|^\beta \textit{ for } T<T_c
\end{equation}
for magnetization and 
\begin{equation}
\label{eq:CV power law}
C_V\propto\left|T_c-T\right|^\alpha
\end{equation}
for specific heat, where $T$ is temperature and $T_c$ is the critical temperature. Each respective least squares power law fit of the resulting magnetization and specific heat data was used to determine the critical temperature of each model and its dependence on the $\tau$ parameter. However, because of limitations in computational power the power law fits applied to the two dimensional Ising model results produced a critical temperature which differed from the analytical critical temperature by 0.061 $k_BT/\epsilon$. Therefore, for all critical temperatures resulting from the power law fits, the difference between the analytical and determined Ising model critical temperature was added to all errors given by the covariance matrix of the fit. However, since the covariance matrix error in the critical temperture was several orders of magnitude smaller than the critical temperature difference error, the critical temperature difference error dominates. 

\subsection{One Dimensional Results\label{sec:1D results}}

\begin{figure*}[!htdp]
\begin{center}
\begin{subfigure}[htdp]{1.0\textwidth} 
\begin{subfigure}[htdp]{0.5\textwidth}
        \centering
        \includegraphics[width=3.0in]{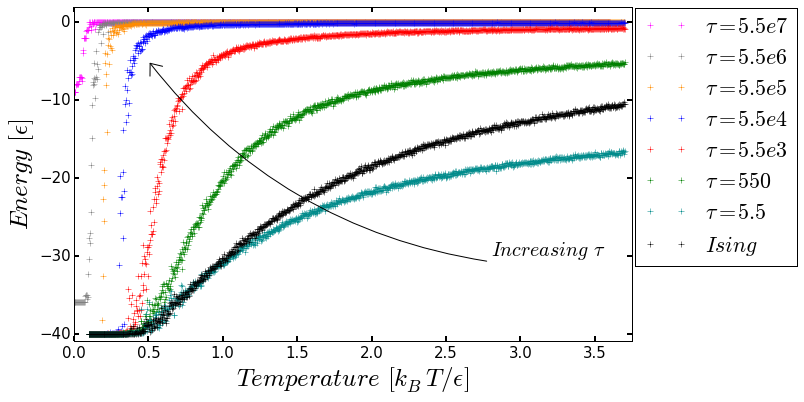}
       \caption{Model 1A energy results.}
\label{fig:1D Bell Energy}
    \end{subfigure}%
 \begin{subfigure}[htdp]{0.5\textwidth}
        \centering
        \includegraphics[width=3.0in]{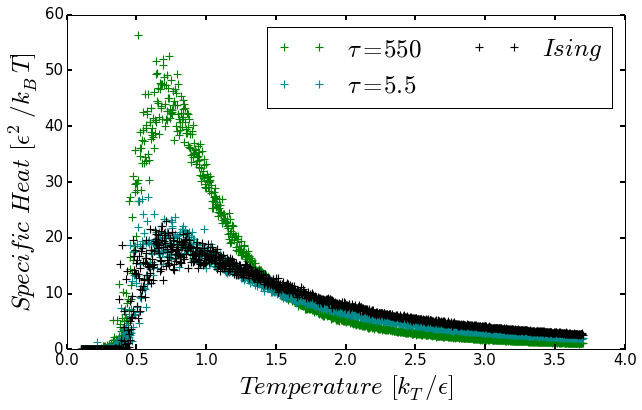}
        \caption{Model 1A specific heat results for low $\tau$ values.}
\label{fig:1D Bell CV1}
    \end{subfigure}%
\end{subfigure}\\
 \begin{subfigure}[htdp]{1.0\textwidth}
 \begin{subfigure}[htdp]{0.5\textwidth}
        \centering
        \includegraphics[width=3.0in]{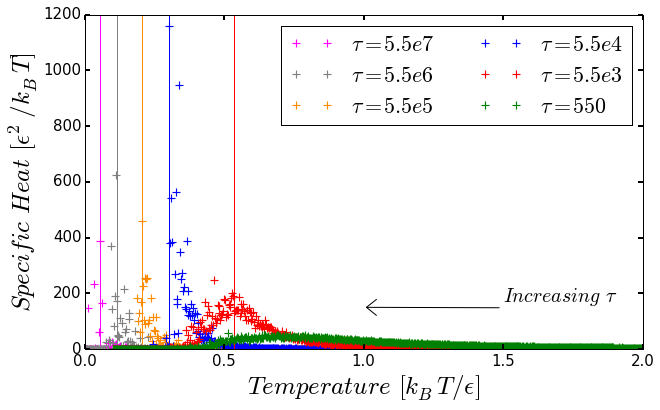}
       \caption{Model 1A specific heat results for high $\tau$ values.}
\label{fig:1D Bell CV2}
    \end{subfigure}%
 \begin{subfigure}[htdp]{0.5\textwidth}
        \centering
        \includegraphics[width=3.0in]{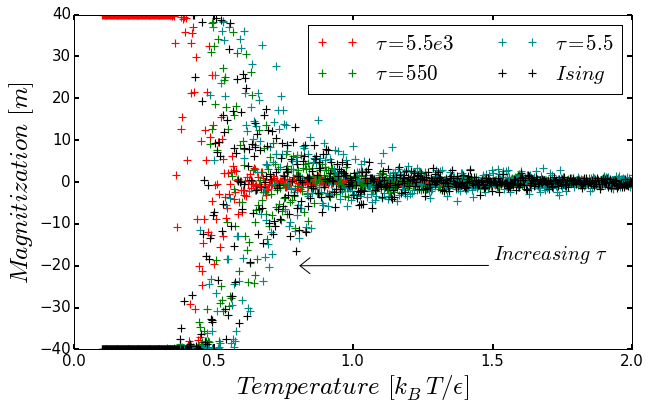}
        \caption{Model 1A magnetization results for low $\tau$ values.}
\label{fig:1D Bell M1}
    \end{subfigure}%
\end{subfigure}
 \begin{subfigure}[htdp]{1.0\textwidth}
 \begin{subfigure}[htdp]{0.5\textwidth}
        \centering
        \includegraphics[width=3.0in]{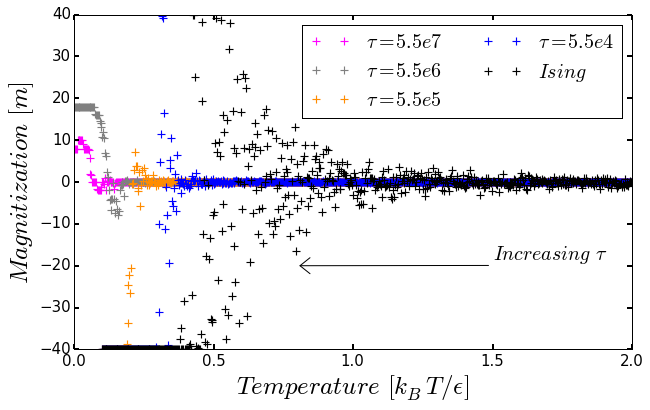}
       \caption{Model 1A magnetization results for high $\tau$ values.}
\label{fig:1D Bell M2}
    \end{subfigure}%
 \begin{subfigure}[htdp]{0.5\textwidth}
        \centering
        \includegraphics[width=3.0in]{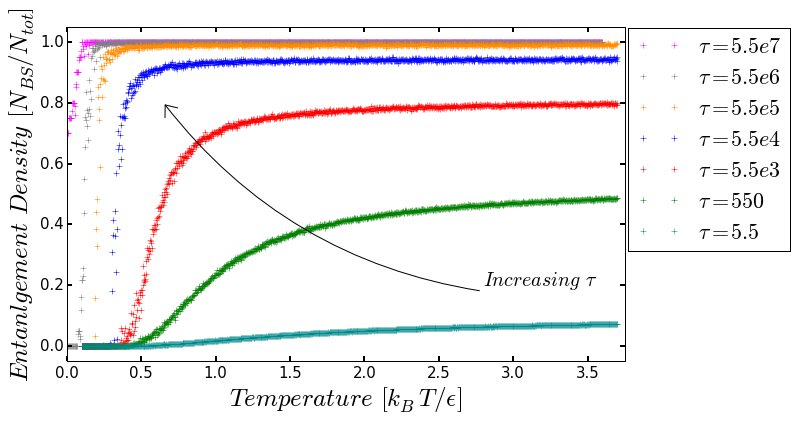}
        \caption{Model 1A entanglement density results.}
\label{fig:1D Bell ED}
    \end{subfigure}%
\end{subfigure}
\caption[Model 1A Results]{Results for Model 1A. 1D Ising Model results are also shown for comparison. Direction in which the thermodynamical results change with increasing $\tau$ values (decoherence time) are also indicated. Vertical lines indicate the critical temperatures determined from the fit of both the specific heat and magnetization. As $\tau$ increases, the critical temperatures reflected in these thermodynamical results decrease. Fig.~\ref{fig:1D Bell Energy}: for low $\tau$ values the resulting energies resemble that of the 1D Ising Model. As the $\tau$ parameter increases, the temperature dependent energy results approach a step function. Fig.~\ref{fig:1D Bell CV1}: the specific heats functionally resemble those of the 1D Ising Model, with the lower $\tau$ value resembling the 1D Ising model very closely. Fig.~\ref{fig:1D Bell CV2}: as $\tau$ increases the critical temperature decreases while the specific heat curves become steeper and begin to resemble a delta function. Fig.~\ref{fig:1D Bell M1}: as $\tau$ increases, the temperature at which the system gains an average non-zero magnetization decreases. No definite critical transition temperature is apparent. Fig.~\ref{fig:1D Bell M2}: as $\tau$ increases, the temperature at which the system gains an average non-zero magnetization decreases. Also, with increasing $\tau$ a definite critical transition temperature becomes more apparent, where it was not for lower $\tau$ values. Fig.~\ref{fig:1D Bell ED}: as $\tau$ is increased, the maximum entanglement density of the system increases. Also, with increasing $\tau$, the temperature at which the system transitions to a near zero entanglement density decreases and the temperature dependent entanglement density approaches a step function.}
\label{fig:1D Bell Results}
\end{center}
\end{figure*}

\begin{figure*}[!htdp]
\begin{center}
\begin{subfigure}[htdp]{1.0\textwidth} 
\begin{subfigure}[htdp]{0.5\textwidth}
        \centering
        \includegraphics[width=3.0in]{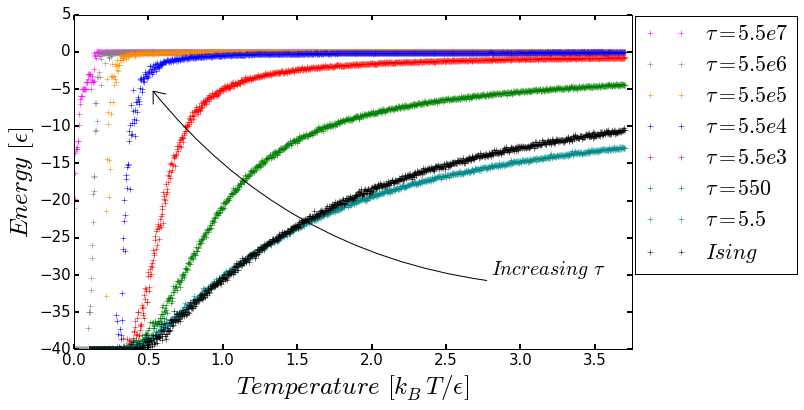}
       \caption{Model 1B energy results.}
\label{fig:1D IB Energy}
    \end{subfigure}%
 \begin{subfigure}[htdp]{0.5\textwidth}
        \centering
        \includegraphics[width=3.0in]{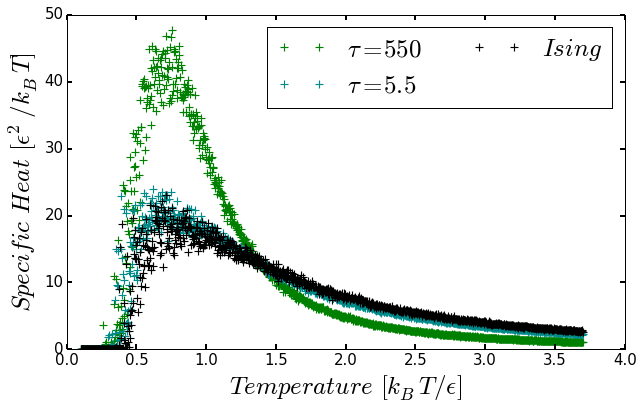}
        \caption{Model 1B specific heat results for low $\tau$ values.}
\label{fig:1D IB CV1}
    \end{subfigure}%
\end{subfigure}\\
 \begin{subfigure}[htdp]{1.0\textwidth}
 \begin{subfigure}[htdp]{0.5\textwidth}
        \centering
        \includegraphics[width=3.0in]{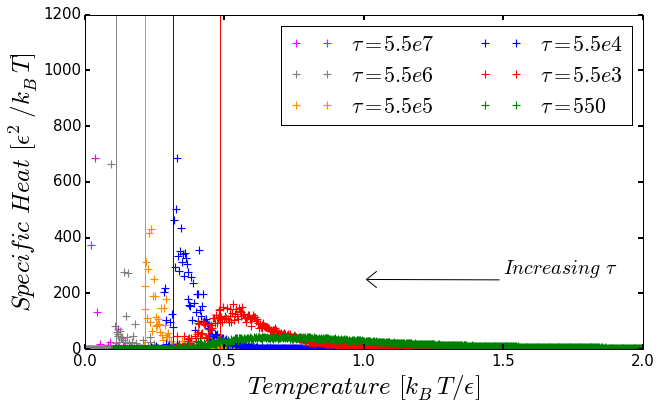}
       \caption{Model 1B specific heat results for high $\tau$ values.}
\label{fig:1D IB CV2}
    \end{subfigure}%
 \begin{subfigure}[htdp]{0.5\textwidth}
        \centering
        \includegraphics[width=3.0in]{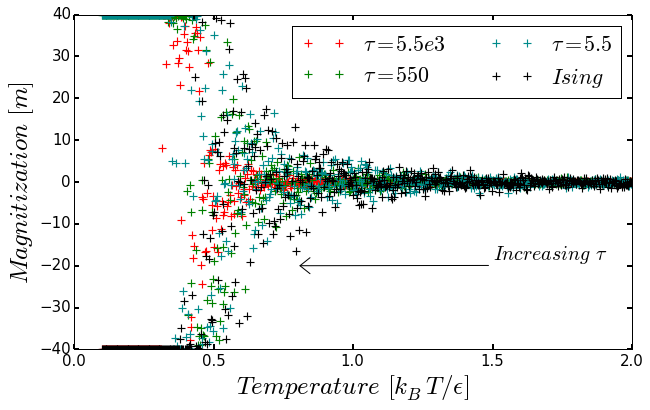}
        \caption{Model 1B magnetization results for low $\tau$ values.}
\label{fig:1D IB M1}
    \end{subfigure}%
\end{subfigure}
 \begin{subfigure}[htdp]{1.0\textwidth}
 \begin{subfigure}[htdp]{0.5\textwidth}
        \centering
        \includegraphics[width=3.0in]{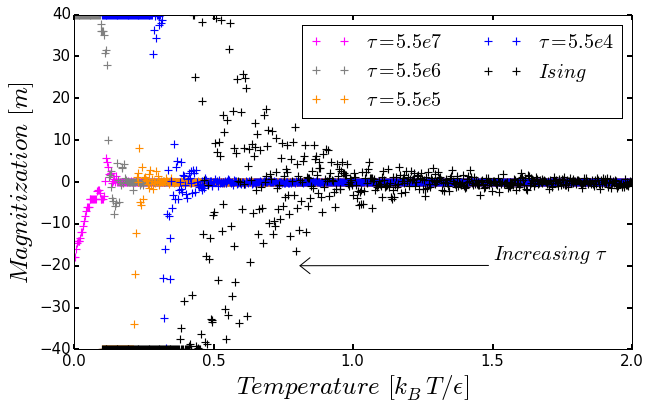}
       \caption{Model 1B magnetization results for high $\tau$ values.}
\label{fig:1D IB M2}
    \end{subfigure}%
 \begin{subfigure}[htdp]{0.5\textwidth}
        \centering
        \includegraphics[width=3.0in]{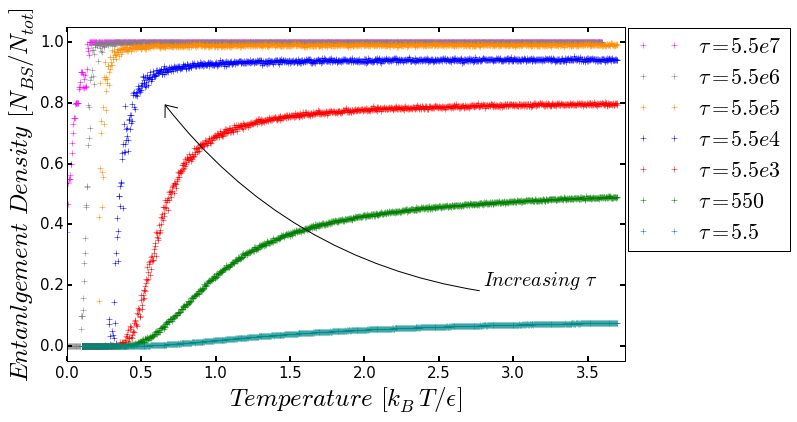}
        \caption{Model 1B entanglement density results.}
\label{fig:1D IB ED}
    \end{subfigure}%
\end{subfigure}
\caption[Model 1B Results]{Results for Model 1B. 1D Ising Model results are also shown for comparison. Direction in which the thermodynamical results change with increasing $\tau$ values (decoherence time) are also indicated. Vertical lines indicate the critical temperatures determined from the fit of both the specific heat and magnetization.As $\tau$ increases, the critical temperatures reflected in these thermodynamical results decrease. Fig.~\ref{fig:1D IB Energy}: for low $\tau$ values the resulting energies resemble that of the 1D Ising Model. As the $\tau$ parameter increases, the temperature dependent energy results approach a step function. Fig.~\ref{fig:1D IB CV1}: these specific heats functionally resemble those of the 1D Ising Model, with the lower $\tau$ value resembling the 1D Ising model very closely. Fig.~\ref{fig:1D IB CV2}: as $\tau$ increases the critical temperature decreases while the specific heat curves become steeper and begin to resemble a delta function. Fig.~\ref{fig:1D IB M1}: as $\tau$ increases, the temperature at which the system gains an average non-zero magnetization decreases. No definite critical transition temperature is apparent. Fig.~\ref{fig:1D IB M2}: as $\tau$ increases, the temperature at which the system gains an average non-zero magnetization decreases. Also, with increasing $\tau$ a definite critical transition temperature becomes more apparent, where it was not for lower $\tau$ values. Fig.~\ref{fig:1D IB ED}:  As $\tau$ is increased, the maximum entanglement density of the system increases. Also, with increasing $\tau$, the temperature at which the system transitions to a near zero entanglement density decreases and the temperature dependent entanglement density approaches a step function.}
\label{fig:1D Ising and Bell Results}
\end{center}
\end{figure*}

The thermodynamical results (energy, specific heat, magnetization, and entanglement density) for Model 1A are shown in Fig.~\ref{fig:1D Bell Results} and the results for Model 1B are shown in Fig.~\ref{fig:1D Ising and Bell Results}. For low $\tau$ values (approximately $\leq550$ iterations) no critical behavior is displayed by either interaction and the energy and specific heat follow a functionality similar to the  one-dimensional Ising model. In order to determine the variation of each interaction from the 1D Ising Model, the energy and specific heat  data of Models 1A and 1B with the $\tau$ values of 5.5 and 550 iterations as well as the 1D Ising Model were fit to the known one-dimensional Ising model energy, 
\begin{equation}
-N\epsilon\tanh{\frac{\epsilon}{k_BT}},
\label{eq:1D Ising energy}
\end{equation} 
and specific heat,
\begin{equation}
\frac{N\epsilon^2}{k_BT^2}\sech{\frac{\epsilon}{k_BT}}
\label{eq:1D Ising CV}
\end{equation}
where $N$ is the particle number or lattice length of the system. The varying lattice lengths which resulted from the fits are given in Table~\ref{table:N Fits}. As expected the 1D Ising Model fits give the correct particle number of 40. However, both the fits of Model 1A and Model 1B gave incorrect lattice lengths indicating that even at low $\tau$ values the Bell state and decoherence interactions alter the thermodynamics of the system. 

\begin{table}[H]
\caption{Resulting lattice lengths from fits of low $\tau$ value energies and specific heats to Eqn.~\ref{eq:1D Ising energy} and~\ref{eq:1D Ising CV} for the 1D Ising Model, Model 1A, and Model 1B. Reported lengths are given in number of particles in lattice. All fitted models had an actual lattice length of 40 particles.\label{table:N Fits}}
\begin{ruledtabular}
\begin{tabular}{|c|c|m{1.8cm}|m{1.8cm}|}
$\tau$ & 1D Ising Model & Model 1A & Model 1B\\
\hline
None & $40.0\pm0.4$ & - & -\\
5.5 & - & $31.1\pm0.1$ & $35.4\pm0.1$ \\
550 & - & $45.1\pm0.7$ & $47.7\pm0.4$ \\
\end{tabular}
\end{ruledtabular}
\end{table}

\begin{center}
\begin{figure}[H]
\includegraphics[width=3.25in]{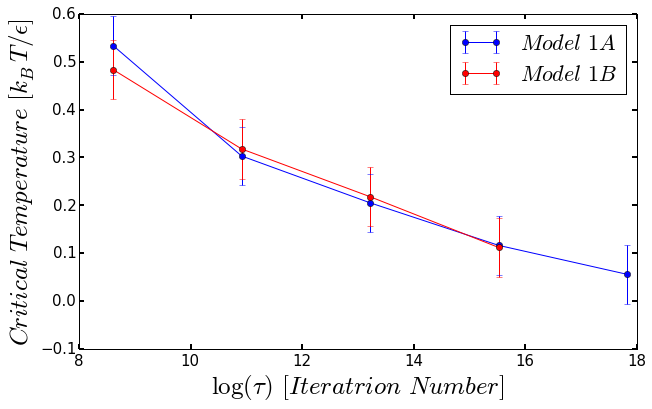}%
\caption{One dimensional critical temperatures vs. $\log(\tau)$ for relevant $\tau$ values. \label{fig:1D Crit Temps}}
\end{figure}
\end{center}

For larger $\tau$ values ($\tau>550$ iterations), corresponding to longer decoherence times, both Model 1A and Model 1B began to display critical behavior. Thus the magnetization and specific heat data for these higher $\tau$ values were fit to the respective power laws given by Eqn.~\ref{eq:M power law} and Eqn.~\ref{eq:CV power law}.. For very large $\tau$ values ($\tau\geq I_{tot}$), the system becomes completely or almost completely saturated with Bell states resulting in all thermodynamical quantities remaining relatively constant even at low temperatures. This saturation prevented a critical temperature from being determined for Models 1A and 1B with $\tau=I_{tot}$. A plot of the critical temperature vs. the natural logarithm of $\tau$ for both Model 1A and Model 1B is shown in Fig.~\ref{fig:1D Crit Temps}. As can be seen in Fig.~\ref{fig:1D Crit Temps}, the critical temperature for both models decreases at about the same rate as the $\tau$ parameter increases. In addition to this, as the $\tau$ parameter increases the energy, magnetization, and entanglement density of both models converge to a step function, where the step occurs at the critical temperature. This in turn causes the specific heat to converge functionally to a delta spike. 

\FloatBarrier

\subsection{Two Dimensional Results\label{sec:2D results}}

\begin{figure*}
\begin{center}
\begin{subfigure}[htdp]{1.0\textwidth} 
\begin{subfigure}[htdp]{0.5\textwidth}
        \centering
        \includegraphics[width=3.0in]{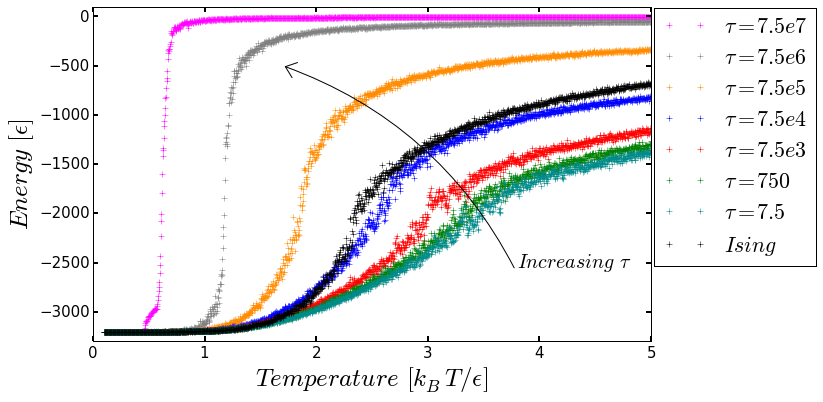}
       \caption{Model 2A energy results.}
\label{fig:2D Bell Energy}
    \end{subfigure}%
 \begin{subfigure}[htdp]{0.5\textwidth}
        \centering
        \includegraphics[width=3.0in]{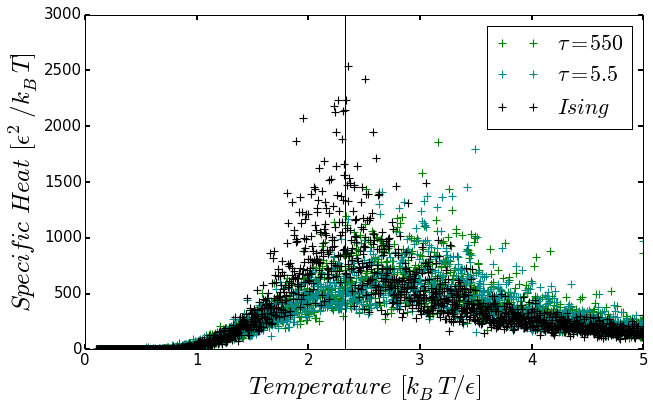}
        \caption{Model 2A specific heat results for low $\tau$ values.}
\label{fig:2D Bell CV1}
    \end{subfigure}%
\end{subfigure}\\
 \begin{subfigure}[htdp]{1.0\textwidth}
 \begin{subfigure}[htdp]{0.5\textwidth}
        \centering
        \includegraphics[width=3.0in]{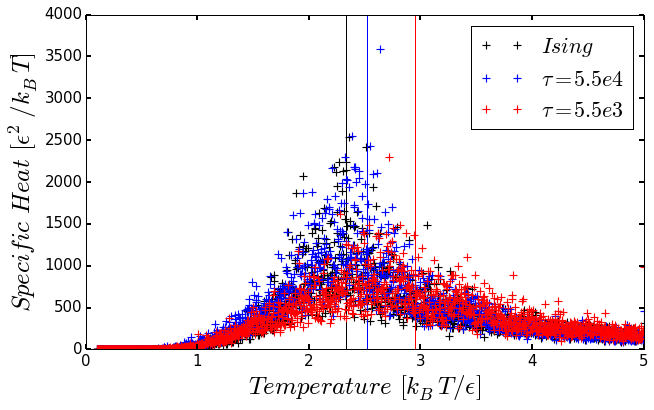}
       \caption{Model 2A specific heat results for mid-range $\tau$ values.}
\label{fig:2D Bell CV2}
    \end{subfigure}%
 \begin{subfigure}[htdp]{0.5\textwidth}
        \centering
        \includegraphics[width=3.0in]{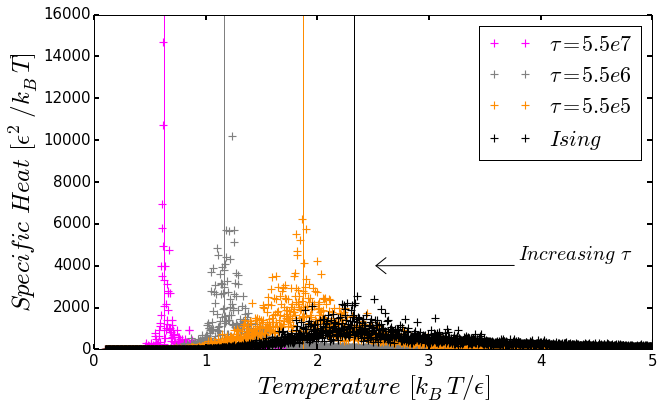}
        \caption{Model 2A specific heat results for high $\tau$ values.}
\label{fig:2D Bell CV3}
    \end{subfigure}%
\end{subfigure}
 \begin{subfigure}[htdp]{1.0\textwidth}
 \begin{subfigure}[htdp]{0.5\textwidth}
        \centering
        \includegraphics[width=3.0in]{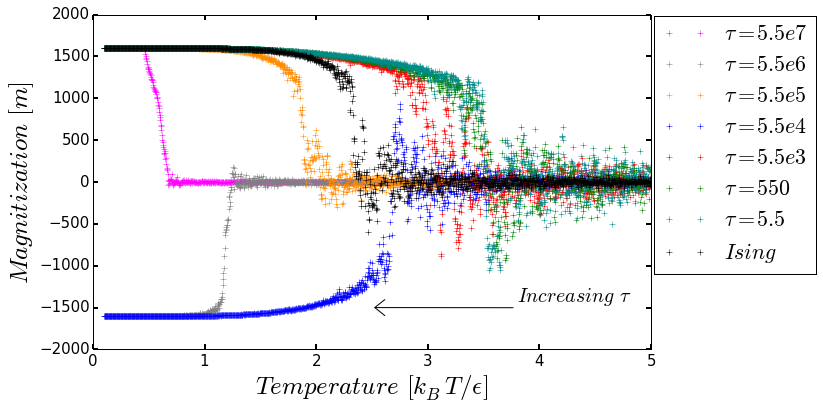}
       \caption{Model 2A magnetization results.}
\label{fig:2D Bell M}
    \end{subfigure}%
 \begin{subfigure}[htdp]{0.5\textwidth}
        \centering
        \includegraphics[width=3.0in]{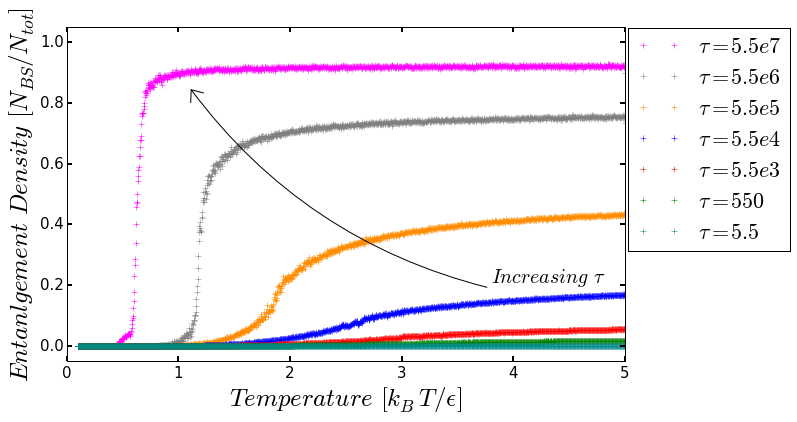}
        \caption{Model 2A entanglement density results.}
\label{fig:2D Bell ED}
    \end{subfigure}%
\end{subfigure}
\caption[Model 2A Results]{Results for Model 2A. 2D Ising Model results are also provided for comparison.  Direction in which the thermodynamical results change with increasing $\tau$ values (decoherence time) are also indicated. Vertical lines indicate the critical temperatures determined from the fit of both the specific heat and magnetization.As $\tau$ increases, the critical temperatures reflected in these thermodynamical results decrease, diverging from the 2D Ising Model. Fig.~\ref{fig:2D Bell Energy}: at low $\tau$ values, the energy functions are lower and smoother than the 2D Ising model energy. As $\tau$ increases, the energy functions become more drastic surpassing the 2D Ising model energy and approaching a step function. Fig.~\ref{fig:2D Bell CV1}: at these low $\tau$ values the specific heat results are relatively smooth displaying no apparent critical temperature, in contrast to the 2D Ising Model also shown. Fig.~\ref{fig:2D Bell CV2}: for these $\tau$ values the specific heats display critical behavior similar to the 2D Ising Model. However, the critical temperatures are lower than that of the 2D Ising Model. Fig.~\ref{fig:2D Bell CV3}: at these higher $\tau$ values the critical temperatures displayed by the specific heats are lower than that of the 2D Ising Model and become lower with increasing $\tau$. Also, as $\tau$ increases, the specific heat transitions become more drastic and begin to resemble a delta function. Fig.~\ref{fig:2D Bell M}: as the $\tau$ parameter increases, the critical temperature of the magnetic transition decreases. In addition, as $\tau$ increases, the transition becomes more drastic and begins to resemble a step function at very high $\tau$ values. Fig.~\ref{fig:2D Bell ED}: as $\tau$ is increased, the maximum entanglement density of the system increases. Also, with increasing $\tau$, the temperature at which the system transitions to a near zero entanglement density decreases and the temperature dependent entanglement density approaches a step function.}
\label{fig:2D Bell Results}
\end{center}
\end{figure*}

\begin{figure*}
\begin{center}
\begin{subfigure}[htdp]{1.0\textwidth} 
\begin{subfigure}[htdp]{0.5\textwidth}
        \centering
        \includegraphics[width=3.0in]{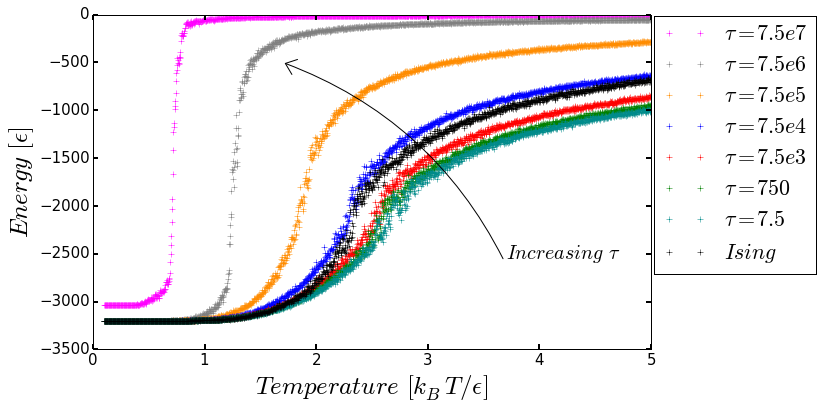}
       \caption{Model 2B energy results.}
\label{fig:2D IB Energy}
    \end{subfigure}%
 \begin{subfigure}[htdp]{0.5\textwidth}
        \centering
        \includegraphics[width=3.0in]{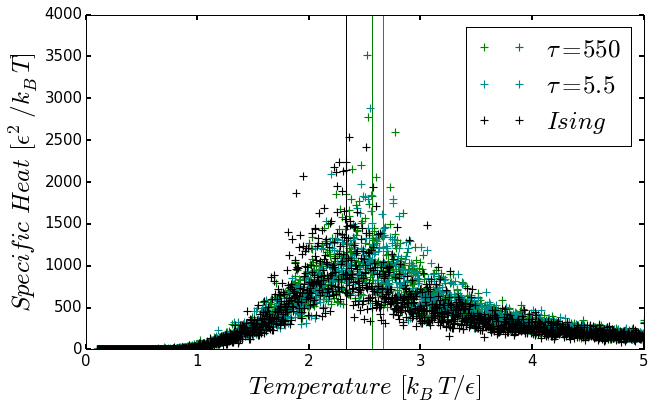}
        \caption{Model 2B specific heat results for low $\tau$ values.}
\label{fig:2D IB CV1}
    \end{subfigure}%
\end{subfigure}\\
 \begin{subfigure}[htdp]{1.0\textwidth}
 \begin{subfigure}[htdp]{0.5\textwidth}
        \centering
        \includegraphics[width=3.0in]{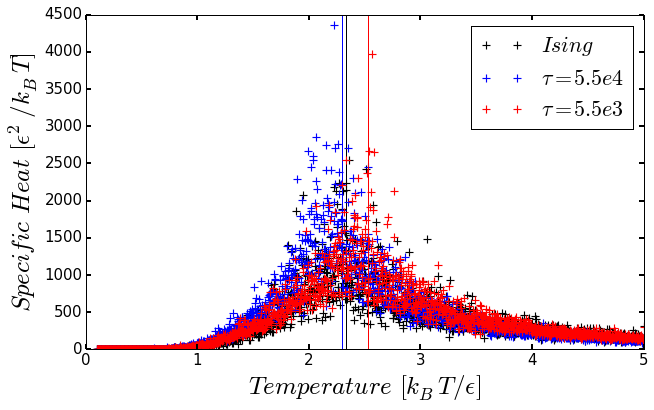}
       \caption{Model 2B specific heat results for mid-range $\tau$ values.}
\label{fig:2D IB CV2}
    \end{subfigure}%
 \begin{subfigure}[htdp]{0.5\textwidth}
        \centering
        \includegraphics[width=3.0in]{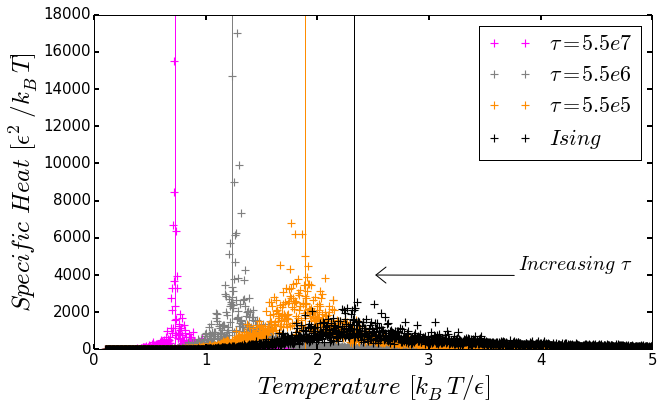}
        \caption{Model 2B specific heat results for high $\tau$ values.}
\label{fig:2D IB CV3}
    \end{subfigure}%
\end{subfigure}
 \begin{subfigure}[htdp]{1.0\textwidth}
 \begin{subfigure}[htdp]{0.5\textwidth}
        \centering
        \includegraphics[width=3.0in]{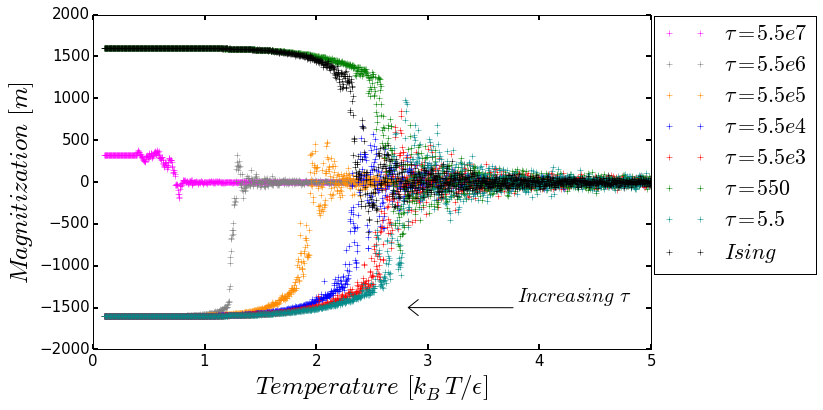}
       \caption{Model 2B magnetization results.}
\label{fig:2D IB M}
    \end{subfigure}%
 \begin{subfigure}[htdp]{0.5\textwidth}
        \centering
        \includegraphics[width=3.0in]{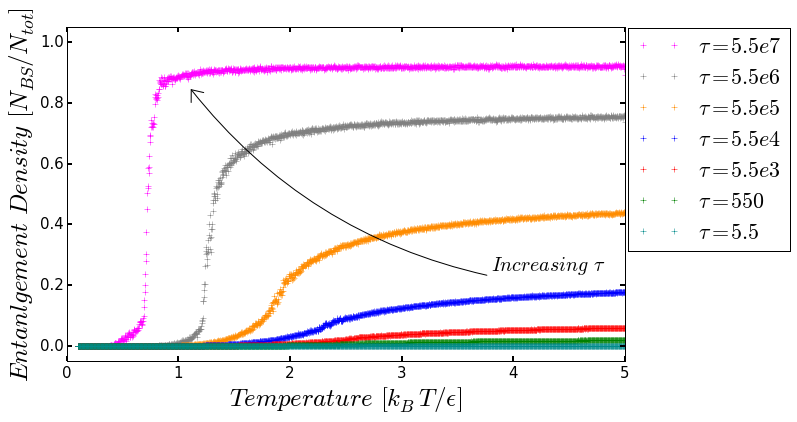}
        \caption{Model 2B entanglement density results.}
\label{fig:2D IB ED}
    \end{subfigure}%
\end{subfigure}
\caption[Model 2B Results]{Results for Model 2B. 2D Ising Model results are also provided for comparison.  Direction in which the thermodynamical results change with increasing $\tau$ values (decoherence time) are also indicated. Vertical lines indicate the critical temperatures determined from the fit of both the specific heat and magnetization.As $\tau$ increases, the critical temperatures reflected in these thermodynamical results decrease, diverging from the 2D Ising Model. Fig.~\ref{fig:2D IB Energy}: At low $\tau$ values, the energy functions are lower but very similar to the 2D Ising model energy. At mid-range $\tau$ values, the energies are very similar and close in magnitude to the 2D Ising model. As $\tau$ increases, the energy functions become more drastic surpassing the 2D Ising model energy and approaching a step function. Fig.~\ref{fig:2D IB CV1}:  for these low $\tau$ values, the specific heats display critical behavior similar to the 2D Ising Model. However, the critical temperatures are lower than that of the 2D Ising Model. Fig.~\ref{fig:2D IB CV2}: for these $\tau$ values the specific heats resemble the 2D Ising Model results. However, the critical temperatures are lower than that of the 2D Ising Model, with the critical temperature for $\tau=5.5e4$ iterations residing very close to the 2D Ising Model critical temperature. Fig.~\ref{fig:2D IB CV3}: at these higher $\tau$ values the critical temperatures displayed by the specific heats are lower than that of the 2D Ising Model and become lower with increasing $\tau$. Also, as $\tau$ increases, the specific heat transitions become more drastic and begin to resemble a delta function. Fig.~\ref{fig:2D IB M}: at lower $\tau$ values the magnetizations resemble that the the 2D Ising Model. As the $\tau$ parameter increases, the critical temperature of the magnetic transition decreases. In addition, as $\tau$ increases, the transition becomes more drastic and begins to resemble a step function at very high $\tau$ values. Fig.~ref{fig:2D IB ED}: as $\tau$ is increased, the maximum entanglement density of the system increases. Also, with increasing $\tau$, the temperature at which the system transitions to a near zero entanglement density decreases and the temperature dependent entanglement density approaches a step function.}
\label{fig:2D Ising and Bell Results}
\end{center}
\end{figure*}

The thermodynamical results (energy, specific heat, magnetization, and entanglement density) for Model 2A are shown in Fig.~\ref{fig:2D Bell Results} and the results for Model 2B are shown in Fig.~\ref{fig:2D Ising and Bell Results}. For low $\tau$ values ($\tau\leq 750$ iterations) and short decoherence tomes, Model 2Al did not show any apparent critical behavior where Model 2Bl did. Therefore, where critical behavior was apparent a least squares fit of the magnetization and specific heat to the respective power laws,  Eqn.~\ref{eq:M power law} and~\ref{eq:CV power law}, were applied to determine the critical temperature. Just as with the one-dimensional models, as the $\tau$ parameter increases the energy, magnetization, and entanglement density of both models converge to a step function and the specific heat converges to a delta spike. In addition, large $\tau$ values cause the system to become saturated with Bell states resulting in a loss of critical behavior. 

\begin{center}
\begin{figure}[H]
\includegraphics[width=3.25in]{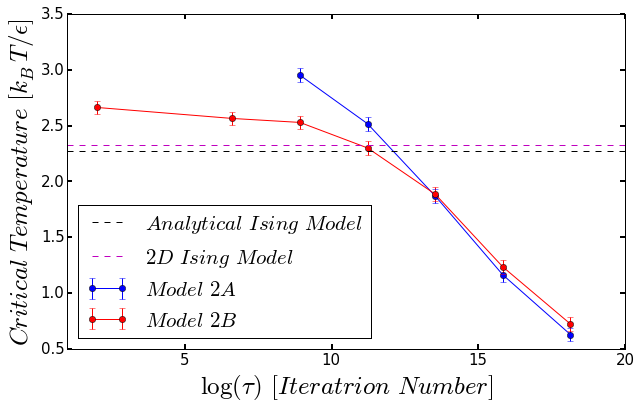}%
\caption{Two dimensional critical temperatures vs. $\log(\tau)$ for relevant $\tau$ values. Both the accepted critical temperature for the two dimensional Ising model and that given by the model are shown as well. \label{fig:2D Crit Temps}}
\end{figure}
\end{center}

A plot of the critical temperature vs. the natural logarithm of the $\tau$ parameter for both Model 2A and Model 2B are shown in Fig.~\ref{fig:2D Crit Temps}. In contrast to the one-dimensional case, $T_c$ vs. $\log(\tau)$ for Model 2B begins to diverge from the Model 2A as $\tau$ becomes smaller, approaching a constant value. Also, this constant value differs from the critical temperature of the pure Ising model.  

\FloatBarrier

\section{Conclusion/Future Work}

By utilizing an Ising-inspired numerical model, the temperature dependent effects of spontaneous quantum teleportation and pair swapping on several  thermodynamical quantities was examined within a one- and two-dimensional spin lattice. Several models were developed by modifying the pure Ising model spin lattice to include a Bell state projection interaction and quantum decoherece interaction instead of and in addition to the typical spin flip of the Ising model. By executing each model via the Metropolis algorithm the temperature dependent effects of the interactions on multiple thermodynamical quantities was determined. In addition, the time dependent decoherence parameter, $\tau$, was also varied to determine its effect on the thermodynamical quantities. The resulting thermodynamical quantities of energy, specific heat, and magnetization were compared with those of the pure Ising model in both one and two dimensions. 

In one dimension, Model 1A and Model 1B, the preliminary results at low $\tau$ values show no critical behavior and follow those of the Ising model. However, as the $\tau$ parameter is increased both Model 1A and Model 1B developed critical behavior in the thermodynamical quantities, something not seen in the one-dimensional Ising model. However, no difference in the $\tau$ dependent critical temperatures between Model 1A and Model 1B was observed. 

The preliminary two-dimensional results, Model 2A and Model 2B, displayed a variation between themselves as well as with the pure two-dimensional Ising model. Model 2A displayed no apparent critical behavior at low $\tau$ values where as the 2D Ising model and Model 2B did. In addition, at these low $\tau$ values the critical temperatures of Model 2B were higher than that of the pure two-dimensional Ising model. As the $\tau$ parameter was increased, both Model 2A and Model 2B experience critical behavior. For moderate $\tau$ values, the critical temperatures of Model 2A and Model 2B were very near that of the pure two-dimensional Ising model. However, as $\tau$ was increased these critical temperatures become much lower than the critical temperature of the pure two-dimensional Ising model. For high $\tau$ values, the critical temperatures of Model 2A and Model 2B are nearly identical, however as $\tau$ is decreased the critical temperatures of the two models diverge, with the critical temperature of Model 2B approaching a constant value.  

Future work investigating the thermodynamical effects of spontaneous quantum teleportation in a spin lattice will include working to determine an analytical Hamiltonian and partition function for all models. If an analytical result can be determined it can be compared with those given by the numerical models. Future work on the numerical model will include the addition of dynamic motion of the spins within the lattice. This dynamic motion should allow more long range quantum teleportaion and quantum decoherence effects to take place. In addition, both models will be expanded to include more general superimposed spin states instead of the binary spin states currently used. These superimposed spin states can be modeled by the spin vector lying on the  surface of a Bloch sphere. Quantum decoherence could then be represented as the decay of the spin vector from the surface of the Bloch sphere. 

\begin{acknowledgments}
We would like to thank Karl Saunders, Ryan Morshead, and Matthew Moelter for stimulating conversations and valuable feedback.
\end{acknowledgments}

\onecolumngrid
\twocolumngrid

\onecolumngrid
\twocolumngrid

\onecolumngrid
\twocolumngrid

\onecolumngrid
\twocolumngrid

\FloatBarrier


\begin{thebibliography}{99}

\bibitem{Qteleport}C. H. Bennett {\it et al.}, Phys. Rev. Letters {\bf70}, 1895 (1993).

\bibitem{ExpQtel}D. Bouwmeester {\it et al.}, Nature {\bf390} (1997).

\bibitem{SolidStateQtel}W. Pfaff {\it et al.}, Science {\bf345}, 6196 (2014).

\bibitem{Exp Enganglement}B. Julsgaard, A. Kozhekin, E. S. Polzik, Nature {\bf413} (2001).

\bibitem{Qkey protocol}N. Zhou {\it et al.}, Optics Communications {\bf284} (2011).

\bibitem{Qkey distribution}H. Lo, M. Curty, B. Qi, Phys. Rev. Letters {\bf108}, 130503 (2012).

\bibitem{Qcomputing}F. Verstraete, J. I. Cirac, Phys. Rev. A {\bf70},  060302(R) (2004)

\bibitem{EPR}A. Einstein, B. Podolsky, and N. Rosen, Phys. Rev. {\bf 47}, 777 (1935).

\bibitem{PairSwap}S. Bose, V. Vedral, P. L. Knight, Phys. Rev. A. {\bf 57}, 2 (1998).

\bibitem{IsingKramers}H. A. Kramers, G. H. Wannier, Phys. Rev. {\bf 60}, 252 (1941).

\bibitem{IsingOnsager}L. Onsager, Phys. Rev. {\bf 65}, 117 (1944).

\bibitem{Metropolis}N. Metropolis {\it et al.}, The Journal of Chemical Physics {\bf 21}, 1087 (1953).

\bibitem{Decoherence}W. H. Zurek, \textit{Decoherence and the Transition from the Quantum to Classical - Revisited}, (Los Alamos Science, 2002).

\end{thebibliography}
\end{document}